\documentclass{aastex}
\usepackage{emulateapj5}
\usepackage{graphicx}
\usepackage{onecolfloat}

\newcommand{\lya}{Ly-$\alpha$ }

\begin{document}
\submitted{draft version; \today}

\shortauthors{TASITSIOMI}
\shorttitle{Line Radiative Transfer in Mesh Simulations}
\twocolumn[%
\title{On the Transfer of Resonant-Line Radiation in mesh simulations}

\author{Argyro Tasitsiomi\altaffilmark{1}}
\vspace{2mm}
\begin{abstract}
The last decade has seen applications of Adaptive Mesh Refinement (AMR) methods for a wide range of 
problems from space physics to cosmology. With the advent of these methods, in which space is discretized into a mesh of many individual cubic elements, the 
contemporary analog of the extensively studied 
line radiative transfer (RT) in a semi-infinite slab  is that of RT in a cube. 
In this study 
we  provide an 
approximate solution of the RT equation, as well as analytic expressions for the probability distribution functions (pdfs) of the properties  of photons emerging from a cube,  and compare them with the corresponding slab problem. These pdfs can be used to perform fast resonant-line RT  in optically thick AMR cells where, otherwise, it could take unrealistically long times to transfer even a handful of photons.
 
\end{abstract}


\keywords{line: formation --- radiative transfer --- resonant}] 

\altaffiltext{1}{Lyman Spitzer Fellow, Department of Astrophysical Sciences,
       Princeton University, Peyton Hall, Ivy Lane, Princeton,  NJ 08544-1001;
       {\tt iro@astro.princeton.edu}}

\section{Introduction}
\label{sec:intro}
The classic problem of resonant radiative transfer (RT)
in a semi-infinite slab has been extensively studied in literature \citep[e.g.,][]{harrington73,neufeld90}.
On the other hand, the last decade has seen applications of Adaptive Mesh Refinement (AMR) methods to problems as diverse as 
solar physics, supernovae and nucleosynthesis, interstellar medium physics, star formation, astrophysical jets, cosmology, etc. \cite[for a summary see, e.g.,][]{norman}.
In mesh-based methods the continuous domain of interest is discretized into a grid of many individual cubic
elements.  
With the advent of  AMR codes, 
the contemporary analogue -- at least in terms of usefulness and applicability range -- of the extensively studied problem of resonant-line RT in a 
slab is the relatively unexplored problem of RT in a cube.

Understanding  resonant RT in optically thick cubes is useful in particular because to  perform RT in  AMR simulations one must solve 
numerous  cube RT problems, since each time a photon enters an AMR cell one has a new cube RT problem. Furthermore, as is the case, e.g., 
for \lya line RT in cosmological simulations of galaxy formation, 
the high resolution
achieved with AMR codes (along with the cooling of the gas),  
leads to very high scattering optical depths~\cite[see][]{tasitsiomi05b}. To obtain results within realistic times, we need a very fast RT algorithm, faster than
the standard direct Monte Carlo approach in which one follows the detailed scattering of photons in each one of the cells.
A way to obtain this  very fast algorithm is to study a priori the RT in cubes of various physical conditions. 
Using the results of such a study, we can avoid following the detailed photon scattering in each cell. Instead, we can immediately take the photon out of the cell treating thus
the problem
on a per cell rather than on a per scattering basis, thus accelerating the RT scheme considerably.

In this  letter we  discuss  results on the  resonant RT in  a cube, and in comparison with resonant RT in a slab of similar physical conditions.

\section{Description of the problem and definitions}
\label{theproblem}
In what follows we assume slab and cube configurations for the scattering medium. The slab is semi-infinite, namely it is finite in one spatial dimension and
infinite in the other two. 
The scattering medium is uniform in its
 properties and has a central source of  center-of-line photons that scatter resonantly before escaping.  
\citet{harrington73} has solved the slab RT equation in the limit of high optical depths and obtained a one parameter solution \citep[also, see][]{neufeld90}.
The parameter is $\alpha \tau_{0}$, with $\alpha=\Delta \nu_{L}/2 \Delta \nu_{D}$ and $\Delta \nu_{L}, \Delta \nu_{D}$ the line natural and thermal Doppler widths, respectively, and
$\tau_{0}$ the center-of-line optical depth from the center of the scattering medium to one of its edges. 
More specifically, 
$\tau_{0}$ is  defined so that
the optical depth at frequency shift $x_{f}=(\nu-\nu_{0})/\Delta \nu_{D}$ is $\tau_{x_{f}}=\tau_{0} \phi(x_{f})$, with $\phi(x_{f})$ the normalized line profile. 
The discussion that
follows applies to optically thick media
($\alpha \tau_{0} > 10^{3}$).
\section{Results}
\label{results}

\subsection{Emerging frequency distribution}
\label{freq}
\subsubsection{ Approximate analytic solution for resonant RT in a cube}
\label{analytic}

Following \citet{harrington73}, one can show that the RT equation we need to solve is 
\begin{equation}
\nabla_{\tau}^{2}J(\vec{\tau};\sigma) +\frac{\partial^{2}J}{\partial \sigma^{2}}=-3 \phi^{2}(x_{f}) \frac{j(\vec{\tau})}{4 \pi} \, 
\label{basic}
\end{equation}
where $J$ is the zeroth moment of the intensity $I$, $\sigma$ is defined through $\partial x_{f}/\partial \sigma= (3/2)^{1/2} \phi(x_{f})$,
$\tau$ is defined through $d\tau_{x_{f}}=\phi(x_{f})d\tau$, $j(\vec{\tau})$ is the emissivity, and $\vec{\tau}$ is measured from the center of the cube. 
This equation is identical to the equation used previously for a semi-infinite slab \citep{unno55,harrington73,neufeld90} or
a spherically symmetric distribution \citep[e.g.,][]{dijkstra05a}. 
The only difference is that  those previous cases  were for one spatial dimension, hence $\nabla_{\tau}^{2}$ consisted of only one term,  
whereas in the case of a cube it has contributions from all three dimensions, i.e. 
$\nabla_{\tau}^{2}=\partial^{2} J/\partial \tau_{x}^{2}+\partial^{2} J/\partial \tau_{y}^{2}+\partial^{2} J/\partial \tau_{z}^{2}$, with
$\tau_{x}, \tau_{y}$ and $\tau_{z}$ the components of $\vec{\tau}$ along the x, y and z-axis, respectively.\footnote{Thus, for $d\tau=-\kappa_{0} ds$, with $s$
measured along the photon propagation direction, and $\kappa_{0}$ such that $\kappa_{0} \phi(x_{f})$ is the 'absorption coefficient', $\tau_{x}, \tau_{y}$ and
$\tau_{z}$ are such that $d\tau_{x}=-\kappa_{0} dx, d\tau_{y}=-\kappa_{0} dy$ and $d\tau_{z}=-\kappa_{0} dz$, respectively. In simple terms, $\tau_{x}, \tau_{y}$ and
$\tau_{z}$ are (modulo a factor $\kappa_{0}$ that depends on the density of the scattering medium)  equivalent to spatial coordinates within the cube.} 

Equation (\ref{basic}) is a linear, inhomogeneous, partial differential equation. To solve it we use the eigenfunction expansion method. 
Namely, motivated by the method of separation of variables (applicable in the case of the corresponding homogeneous equation), we assume that 
the solution can be written as
\begin{equation}
J(\vec{\tau}; \sigma)=\sum_{\alpha,\beta,\gamma=1}^{\infty} X_{\alpha}(\tau_{x}) Y_{\beta}(\tau_{y}) Z_{\gamma}(\tau_{z}) G_{\alpha\beta\gamma}(\sigma) \, .
\label{sum}
\end{equation}
When applying this expansion method in inhomogeneous problems, the idea is that $X_{\alpha},Y_{\beta}$ and $Z_{\gamma}$ will be  known (eigen)functions and $G_{\alpha\beta\gamma}$ 
will be the unknown 
coefficients of the sum (here frequency-dependent)  that are to be determined through the solution process. In reality, we have to specify all four functions since we do not
have any "known" eigenfunctions. We take 
the "known" (eigen)functions of position to be the  solutions
to the "associated" homogeneous ordinary differential equations. The term "associated" here implies that the best choice for the basis position (eigen)functions should be
the solution sets from  Sturm-Liouville problems that closely resemble the problem  being addressed. This will give a set of functions that are orthogonal over the 
domain defined by the problem.  
Thus, to find the position (eigen)functions, we  focus first on the solution of the corresponding homogeneous equation. 
We assume that the solution is separable, namely that  $J(\vec{\tau},\sigma)=R(\vec{\tau})G(\sigma)$. 
Substituting in equation (\ref{basic}) 
and after some rearrangement we get
\begin{equation}
\frac{\nabla_{\tau}^{2}R(\vec{\tau})}{R(\vec{\tau})} =-\frac{1}{G(\sigma)}\frac{\partial^{2} G(\sigma)}{\partial \sigma^{2}}=-\lambda^{2} \, ,
\end{equation}
where we have performed a first separation with separation constant $-\lambda^{2}$.
As already implied by equation (\ref{sum}), we  furthermore assume that $R(\vec{\tau})=X(\tau_{x})Y(\tau_{y})Z(\tau_{z})$ in Cartesian coordinates. 
After performing additional separations we end up with equations of the form
\begin{equation}
\frac{1}{X}\frac{\partial^{2}X}{\partial \tau_{x}^{2}}=-l^{2} 
\end{equation}
and similarly for $Y(\tau_{y}), Z(\tau_{z})$ with separation constants $-m^{2}$ and $-n^{2}$, respectively, and 
with $\lambda^{2}=n^{2}+l^{2}+m^{2}$.
The general solution for each one of these equations consists of both sine and cosine terms.  Since we only consider central point
sources, i.e.  $j(\tau)=\delta(\tau_{x}) \delta(\tau_{y}) \delta(\tau_{z})$, each one of the functions $X$,$Y$, and $Z$ must be separately even. 
Thus, we set  $X(\tau_{x})=A \cos(l\tau_{x}), Y(\tau_{y})=B \cos(m\tau_{y})$ and $Z(\tau_{z})=C \cos(n\tau_{z})$.

We calculate the constants $A,B,C$ using boundary conditions that are generated  assuming the Eddington approximation~\cite[see, e.g., p 322 of][]{eddington26}, where
near isotropy is assumed. Given the optical depths we are concern with, the near isotropy assumption should be fairly accurate.
In fact, we adapt the usual two stream approximation to the cube problem. Instead of assuming mild anisotropy in the form 
of two streams (i.e., $I=I_{1}(\tau)$ for $0 \le \theta < \pi/2$ and $I=I_{2}$ for $\pi/2 \le \theta < \pi$), we  take into account all 3 spatial directions and
treat them equivalently. That is, we use an eight-stream approximation, which nevertheless, leads to the 
same conditions as the two stream approximation. Now, however, these conditions refer separately to each one of the three independent directions. 
Thus, we get 
\begin{eqnarray} 
J(\pm \tau_{0},\tau_{y},\tau_{z};\sigma)=
\pm 2H(\pm \tau_{0},\tau_{y},\tau_{z};\sigma)=\mp \frac{2}{3 \phi(x_{f})}\left.\frac{\partial J}{\partial \tau_{x}}\right|_{\tau_{x}=\pm \tau_{0}} 
\label{bc}
\end{eqnarray}
with $H$  the first moment of $I$ 
and  where, in order to obtain the last equality, we 
used the other Eddington condition, $J=3 Tr(K)$, with $K$ the second moment of $I$ with respect to $\vec{n}$, 
 combined with that $\nabla_{\tau} K/ \phi(x_{f})= H$.  
Using  equation (\ref{bc}) we get
for $l_{\alpha}$ the condition \cite[also see, e.g.,][]{neufeld90}
\begin{equation}
l_{\alpha} \tau_{0} \simeq \pi(\alpha-1/2)\left\{1-\frac{2}{3 \phi \tau_{0}} +O\left[\frac{1}{(\phi \tau_{0})^{2}} \right] \right\} \, ,
\label{condition}
\end{equation}
as well as similar conditions for $m_{\beta}$ and $n_{\gamma}$.

Normalizing the cosine solutions in $[-\tau_{0}, \tau_{0}]$ in each direction separately, the solution to the homogeneous equation RT equation takes the
form
\begin{eqnarray} \nonumber
J(\tau_{x},\tau_{y},\tau_{z};\sigma) & = & \sum_{\alpha,\beta,\gamma=1}^{\infty} \cos(l_{\alpha} \tau_{x})\cos(m_{\beta} \tau_{y})\cos(n_{\gamma} \tau_{z}) \\
 & \times & G_{\alpha\beta\gamma}(\sigma)/\tau_{0}^{3/2} \, .
\label{thirteen}
\end{eqnarray}
Now we proceed to calculate the "coefficients" of the sum given in equation (\ref{sum}). Substituting this in equation (\ref{basic}), and after multiplying by $\cos(l_{\alpha}\tau_{x})$$ \cos(m_{\beta} \tau_{y})$$ \cos(n_{\gamma} \tau_{z})/ \tau_{0}^{3/2}$, and integrating over
volume 
  we get
\begin{equation} 
-\lambda^{2}_{\alpha\beta\gamma} G_{\alpha\beta\gamma}+\frac{\partial^{2} G_{\alpha\beta\gamma}}{\partial \sigma^{2}}= 
-\frac{\sqrt{6}}{4 \pi \tau_{0}^{3/2}} \delta(\sigma) Q_{\alpha\beta\gamma}
\label{final}
\end{equation}
with
\begin{eqnarray} \nonumber
Q_{\alpha\beta\gamma}=\int_{-\tau_{0}}^{\tau_{0}}\int_{-\tau_{0}}^{\tau_{0}}\int_{-\tau_{0}}^{\tau_{0}} j(\vec{\tau}) \cos(l_{\alpha}\tau_{x})\cos(m_{\beta}\tau_{y}) \\ 
\cos(n_{\gamma}\tau_{z}) d\tau_{x}d\tau_{y}d\tau_{z} \, .
\end{eqnarray}
For a source of unit strength at the center of the cube we have $j(\vec{\tau})=\delta(\tau_{x})\delta(\tau_{y})\delta(\tau_{z})$, and thus
$Q_{\alpha\beta\gamma}=1$.  
In equation (\ref{final})  we set $3 \phi(x_{f})^{2}= \sqrt{6} \delta (\sigma)$ \cite[see,][]{harrington73}.

Away from $\sigma=0$ equation (\ref{final}) is homogeneous. 
Imposing the boundary condition $J(\vec{r},\pm \infty)=0$ 
that reflects the fact that we do not expect photons with infinite frequency shifts (thus, $G_{\alpha\beta\gamma}$ 
must go to zero for high -- positive or negative -- $\sigma$ values) we get  
\begin{equation}
G_{\alpha\beta\gamma}(\sigma)=De^{-\lambda_{\alpha\beta\gamma}|\sigma|} \, .
\end{equation} 
We obtain the value of the constant $D$ exactly as in \citet{harrington73}.
Plugging all these in equation (\ref{thirteen}) and, 
since we are interested in the overall spectrum of radiation emerging from one side of the cube, say along the z axis, after integrating over $\tau_{x}, \tau_{y}$ and setting
$\tau_{z}=\tau_{0}$ we get
\begin{eqnarray} \nonumber
J(\tau_{0},\sigma)&=&\sum_{\alpha,\beta,\gamma=1}^{\infty}\frac{\sin(l_{\alpha} \tau_{0})}{l_{\alpha}\tau_{0}}\frac{\sin(m_{\beta} \tau_{0})}{m_{\beta}\tau_{0}}\cos(n_{\gamma} \tau_{0})  \\ 
& \times & \frac{\sqrt{6}}{2\pi}  \frac{e^{-\sqrt{(l_{\alpha}\tau_{0})^{2}+(m_{\beta}\tau_{0})^{2}+(n_{\gamma}\tau_{0})^{2}} |\sigma|/\tau_{0}}}{\sqrt{(l_{\alpha}\tau_{0})^{2}+(m_{\beta}\tau_{0})^{2}+(n_{\gamma}\tau_{0})^{2}}} \, .
\label{final2}
\end{eqnarray}
At this last step we also substituted $\lambda^{2}_{\alpha\beta\gamma}$ with $l_{\alpha}^{2}+m_{\beta}^{2}+n_{\gamma}^{2}$.
For a comparison of the spectrum emerging from one side of a cube to that emerging from one of the two 'sides' of a slab we must multiply 
our cube solution  by a factor of 3 so that both solutions have the same normalization.  
This is so because, for the same central source, we expect 1/6 of the photons to emerge from a certain cube side and 
1/2 of the photons to emerge from a certain slab 'side'.

\begin{figure}[htb]
\begin{center}
\epsscale{0.8}
\plotone{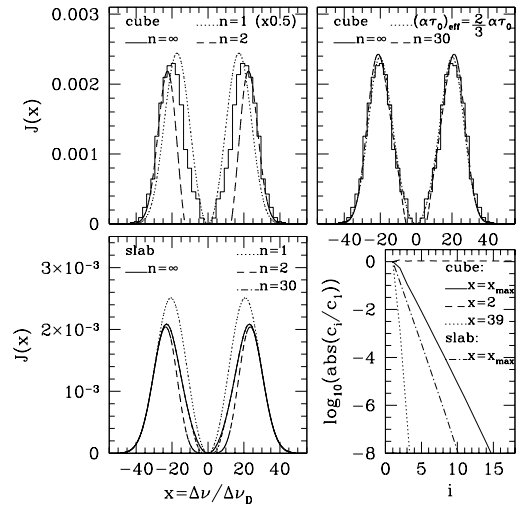}
\caption{{\em Top left panel:} Cube emergent spectrum obtained via
Monte Carlo RT simulation ({\em solid histogram--$n=\infty$}), and approximate analytical spectra  using the one or 
two first terms of the sum given by equation (\ref{final2}) ({\em dotted} and {\em dashed line}, respectively). The $n=1$ spectrum is multiplied by 0.5 to fit within the same
y-axis range as the other spectra. {\em Top right panel:} Same as in left panel but for $n=30$ terms ({\em dashed line}). Also shown
is the spectrum obtained using the slab solution for a smaller, effective $\alpha \tau_{0}$ ({\em dotted line}). {\em Bottom left panel:}
Slab emergent spectra using for the slab infinite number of terms ({\em solid line}--obtained using the closed form solution), one term ({\em dotted line}),
two terms ({\em dashed line}), and the first thirty terms ({\em dot-dashed line}). {\em Bottom right panel:} Logarithm of the absolute value
of the ith term of the sum representing the spectrum for a cube/slab, in units of the first term $c_{1}$. 
Results are shown for the spectrum estimated at different frequency shifts, $x$. For most $x$ values we show results only for the cube, 
since cube and slab
terms exhibit the same convergence behavior, except near the frequencies where they attain their maximum intensity ($x_{max}$).  
All results are for $\alpha \tau_{0}= 2\times 10^{4}$. See text for details.   }
\label{slab_terms}
\end{center}
\end{figure}

Each individual series term for both the cube  and the slab solution \citep[given in][]{harrington73} depends only on $\alpha \tau_{0}$. 
This becomes obvious when one takes into account the definition of $\sigma$ and the 
approximation for $\phi(x_{f}) \simeq \alpha/ \pi x_{f}^{2}$ in the wings, as well as  condition (\ref{condition}).
 The slab solution is an alternate series and can be written in closed form.
The cube solution cannot be written in closed form, but, using
equations (\ref{bc}) and (\ref{condition})  we see that $\sin(l_{\alpha} \tau_{0})\simeq (-1)^{\alpha-1}$ and $\cos(n_{\gamma} \tau_{0})=2 n_{\gamma}/3\phi(x_{f}) (-1)^{\gamma-1}$. Thus, 
the $\sin(l_{\alpha} \tau_{0}) \sin(m_{\beta} \tau_{0}) \cos(n_{\gamma} \tau_{0})$ term in equation (\ref{final2}) can be written as $-(-1)^{\alpha+\beta+\gamma}$,
indicating that the cube series may also be alternating.
Writing this three-sum  series as  one sum, i.e., $\sum_{i=1}^{\infty} c_{i}$, we find that indeed the cube series is alternating as well. Some 
of the first finite sums of the alternating series for a $\alpha \tau_{0}=2 \times 10^{4}$ cube and  slab are shown at the top two panels, and the left bottom panel of Figure \ref{slab_terms}, respectively. 
In the case of 
the cube  the infinite number of terms result (solid histograms)
is obtained by the Monte Carlo method described in detail in \citet{tasitsiomi05b}, whereas for the slab we use the closed form slab analytical solution of \citet{harrington73}.

The cube series solution will be of some practical value, only if a few first terms contribute significantly to the sum. The 
exponential decay in $|\sigma|$ indicates that the
terms should die off for 'high' $\alpha, \beta$ and $\gamma$ values--with the exact values where this happens 
dependent on the frequency (or $\sigma$) one calculates the spectrum at. 
The logarithm of the absolute ratio of the series terms, $c_{i}$, in units of the first term, $c_{1}$, 
for 3 different values of the frequency shift 
is shown in the bottom right panel of Figure \ref{slab_terms}. As before, $\alpha \tau_{0}=2 \times 10^{4}$, but we find that these 'convergence curves' remain identical
for all cube $\alpha \tau_{0}$ at the extremely optically thick regime. The dashed-dotted line is the 'convergence curve' for a slab for 
frequency shift equal to the shift where the emergent spectrum is known to have a maximum \citep[$\sim 0.9 (\alpha \tau_{0})^{1/3}$;][]{harrington73}.
Away from this maximum, the slab convergence curves look identical to those of the cube (dotted and dashed lines).
Clearly, the rate of convergence depends on the frequency the spectrum is calculated at.
Given that the series is alternating, 
with the absolute value of the terms decreasing monotonically (for most $\sigma$ values), the 
absolute error we make by truncating the series at the
ith term  is less or equal to the i+1 term. From the plot, this means that if we only keep the first term, the actual infinite series limit $S$ will be  $\simeq (1\pm 10^{-4})S_{1}$ for
$x_{f}=39$ and $\simeq (1 \pm 0.3) S_{1}$ for $x=x_{max}$, with $S_{i}$ the ith partial sum. We can achieve 
a $3\%$ accuracy at the peak if we go to $i=4$ (i.e., $|S-S_{4}|/S \le 0.03$), whereas for better accuracy we have to exceed $i=30$. Convergence 
becomes extremely slow at values close to the resonance, e.g., $x=2$. Furthermore, 
we find that the up to $i=30$ terms for $x=2$ do not decrease monotonically in the case of the cube -- they do decrease for the slab but extremely slowly.

The usefulness of the slab series solution derived here depends on the application in mind. For example, when studying Ly-$\alpha$ emitters at very high redshifts, it 
is expected that the blue frequencies
will be  absorbed anyway because of  hydrogen intervening between the emitter and the observer, and 
the red wavelengths near the resonance will also be absorbed by the red damping wing. In this case, the poor convergence at frequencies near resonance (and, more generally, 
over the $|x| \le x_{max}$ range) may not be a problem. Clearly, one must decide on the usefulness of the series solution based on the specifics of the application.    

\subsubsection{Approximate cube spectrum in closed form}
\label{cubefromslab}
Based on the similarities between the cube and the slab spectra, 
one might think that the spectrum emerging from one side of a cube may be well described by the slab closed form solution, but for a different,
smaller $\alpha \tau_{0}$ than the actual $\alpha \tau_{0}$ of the cube. Because, for example, when observing the 
spectrum emerging from a cube along the z-direction we lose all photons that in the case of a slab would wander, scatter 
many times along the infinite dimensions until they finally escape from the z-plane. In the case of the cube 
these photons do not contribute to the spectrum we observe from the z-direction because they have already exited the cube
along directions other than z.  
This argument has been already used in \citet{tasitsiomi05b}, where the slab solution was used to describe the spectrum for
a cube by plugging in the slab solution an $\alpha \tau_{0}$  equal to
$2/3$ the actual cube $\alpha \tau_{0}$. This value was motivated by comparing the mean number of scatterings, $N_{sc}$, for photon escape in cubes and slabs.  Since
$N_{sc}$ scales linearly with $\tau_{0}$ for extremely optically thick media \citep{adams72}, by comparing $N_{sc}$ for cubes and slabs one can guess a correct effective $\tau_{0}$ 
(and thus $\alpha \tau_{0}$). 
For the purposes of this previous  study, this effective thickness gave good agreement with simulated spectra for a wide range of physical conditions.  

Attempting a more detailed treatment, we fit cube spectra with the slab solution for the fraction $f$ of $\alpha \tau_{0}$ that must be used in this solution to get the best fit. 
We find that $f$ varies in the $0.66-0.77$ range for $\alpha \tau_{0}$ values in the $2\times 10^{3}-10^{8}$ range. We find
no trend of $f$ with $\alpha \tau_{0}$. 
Within errors, one can use a constant fraction in the above $f$-range regardless of $\alpha \tau_{0}$ since we find that we 
cannot distinguish between the fits obtained when $2/3$ or the exact $f$ value are used.
An example of a fit of the cube spectrum using the slab solution and $f=2/3$ is shown at the top, right panel of
Figure \ref{slab_terms} (the best fit $f$-value for this example is  0.72).
Furthermore, in terms of the spectrum shape (e.g., maximum location and intensity), the above range of $f$ values lead to spectra
indistinguishable by currently existing instruments. 

\subsection{Distribution of exiting direction and point}
\label{exit}
\begin{figure}[thb]
\begin{center}
\epsscale{0.5}
\plotone{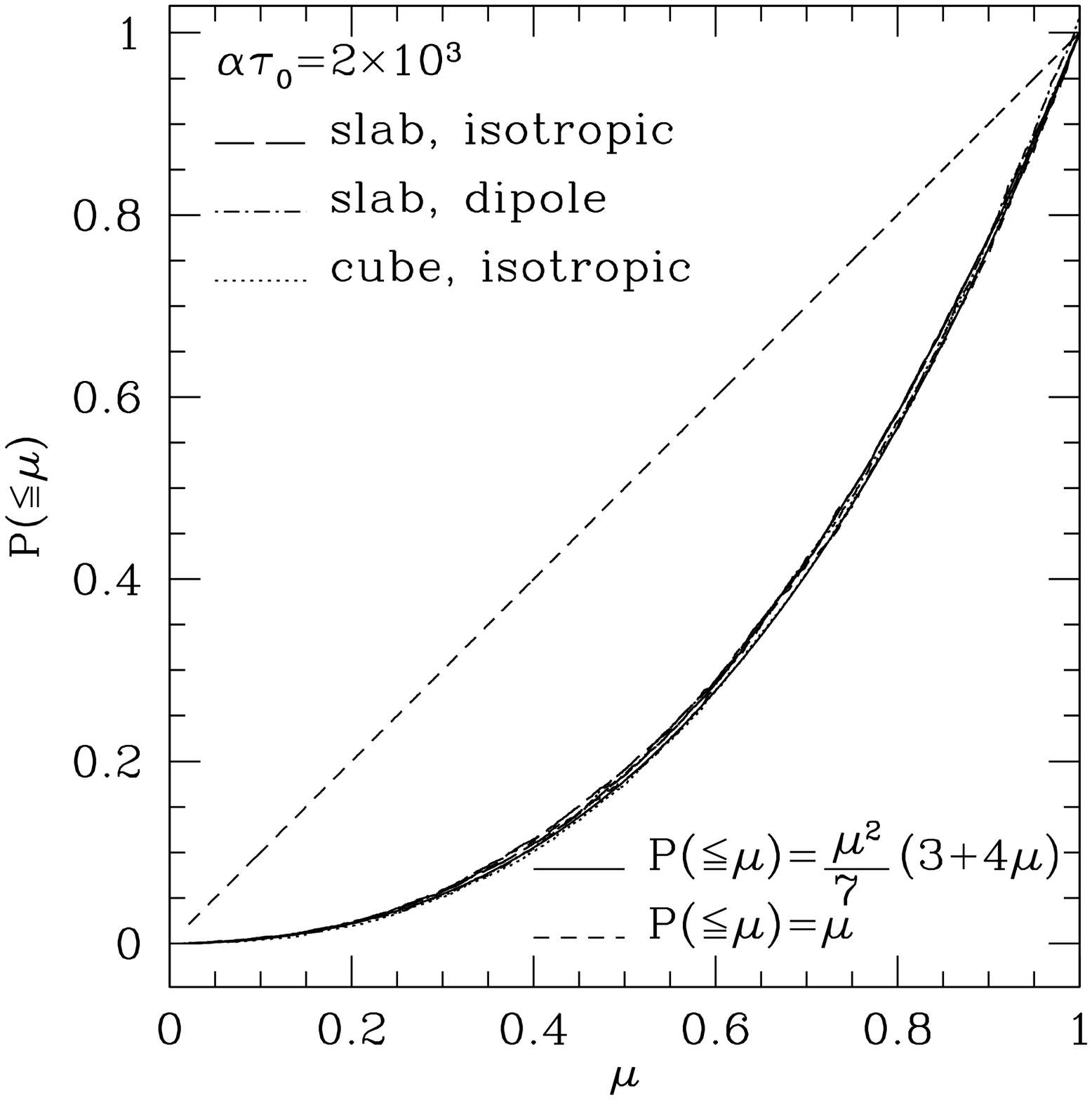}
\plotone{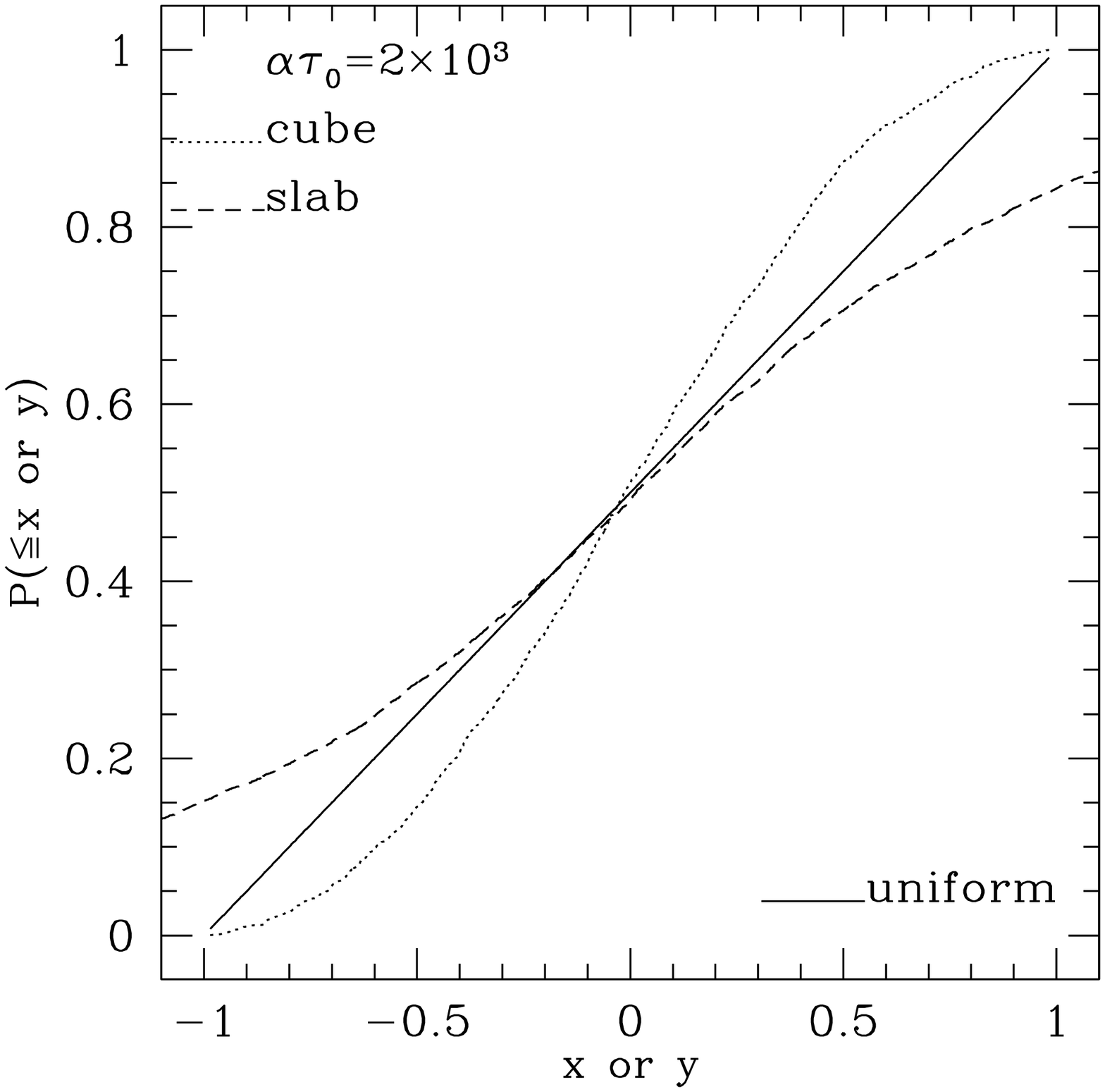}
\caption{{\em Top panel:} Cumulative probability distribution function of the cosine of the angle the photon exiting direction forms
with the normal to the exiting surface ($\mu$). Results are shown for a semi-infinite slab or a cube
({\it long-dashed} and {\it short-dashed lines}, respectively), as well as for an isotropic or dipole  scattering phase function ({\it long-dashed} and
{\it dot-dashed line}, respectively). 
With the {\em solid line}  we show the directionality of radiation emerging from a Thomson-thick electron
medium.
All these distributions are indistinguishable.
Also shown is the case where photons exit 
the slab (or cube) isotropically  ({\it dashed line}). {\em Bottom panel:} Cumulative probability distribution function of exit points of radiation emerging from a slab ({\em dashed line})
and a cube ({\em dotted line}) of the same $\alpha \tau_{0}$. The photons are assumed to emerge along the z direction, hence the distribution refers to either the x or y coordinate,
in units of the size of the cube (or the finite dimension of the slab).}
\label{angle_slab_cube}
\end{center}
\end{figure}

Referring to $\mu$, the cosine of the angle with which the photon is exiting, measured with respect to the normal to the exiting surface,
we show its cumulative probability distribution function (cpdf)  
at the top panel of  Figure \ref{angle_slab_cube}.
We find that this cpdf is the same for a slab or a cube, and
clearly deviates from isotropy (dashed line).
We verify the findings of other studies that in optically thick media photons tend to exit in directions perpendicular to 
the exiting surface \cite[see, e.g.,][]{phillips_meszaros86,ahn_etal02b}. 

In cases of very optically thick media, the emerging radiation directionality approaches the Thomson
scattered radiation emergent from a Thomson-thick electron medium. 
\citet{phillips_meszaros86} found that for a Thomson-thick medium $I(\mu)/I(0)=(1+2\mu)/3$.
Since the number of photons emerging at $\mu$ is $\propto I(\mu) \mu d\mu$, we get
\begin{equation}
P(\le \mu)=\frac{\int_{0}^{\mu} (1+2\mu)\mu d\mu}{\int_{0}^{1} (1+2\mu)\mu d\mu}=\frac{\mu^{2}}{7}(3+4\mu) \, .
\label{dir}
\end{equation}
This dependence is shown in  Figure  \ref{angle_slab_cube} with the solid line. 
It is an excellent description of the directionality of the
spectrum and hence equation (\ref{dir}) can be used to determine the photon exiting direction. 
It has been suggested by some authors \citep{ahn_etal02b} that 
the fact that in optically thick media RT occurs mostly  via wing photons with
the latter being described by a dipole phase function \citep{stenflo80}, and the fact that Thomson scattering  is also described by a 
dipole scattering phase function explain why the resulting $\mu$ cpdfs are similar. 
However, we find the same cpdf 
if we use either an isotropic or a dipole phase distribution. 
For such optical thicknesses the details of the exact phase function do not matter, at least not
with respect to the exiting angle cpdf. 
Regardless of scattering phase function, in such media photons tend to escape along the normal to the exiting surface  where the opacity is smaller.
The results shown are for $\alpha \tau_{0}=2 \times 10^{3}$, but similar distributions are obtained for thicker media.

In the case of the slab the azimuthal angle $\phi$ is distributed uniformly in $[0,2\pi]$.
In the case of the cube there are small deviations from  uniformity. This is expected since
the previous distribution for $\mu$ is found to be valid for {\em all} sides of the cube. In the simplest case where we observe along the
z-axis (in which case the spherical coordinate $\phi$ angle is the actual azimuthal angle we refer to), all direction cosines -- $\cos\theta$ and
$\sin\theta\cos\phi$ and $\sin\theta\sin\phi$ -- follow the distribution given in
equation (\ref{dir}), thus $\phi$ cannot be exactly uniformly distributed. However, the deviations from uniformity are small. 

The distribution of exiting points  for both a cube and a slab are shown at the bottom panel of  Figure \ref{angle_slab_cube}. 
For this figure we assume that we observe photons emerging along the $z$ direction and
we record the $x$ and $y$ coordinates of their exit points (in units of the size of the cube). In the case of 
RT in a cube the distribution is pretty close to uniform. 
In the case of the slab, despite it being semi-infinite, for any practical purposes one can assume that all photons exit at most within
$|x|(|y|) \leq 5$ (not shown in figure). 

\acknowledgements 
I want to thank  A.V. Kravtsov for discussions and comments on the manuscript, and M. Dijkstra, N.Y. Gnedin and D. Neufeld for useful interaction.   
This work was supported, in part, by the Kavli Institute for
Cosmological Physics at The University of Chicago and by the National
Science Foundation under grant NSF PHY 0114422.

\bibliography{ms}

\begin{thebibliography}{11}
\expandafter\ifx\csname natexlab\endcsname\relax\def\natexlab#1{#1}\fi

\bibitem[{{Adams}(1972)}]{adams72}
{Adams}, T.~F. 1972, \apj, 174, 439

\bibitem[{{Ahn} {et~al.}(2002){Ahn}, {Lee}, \& {Lee}}]{ahn_etal02b}
{Ahn}, S., {Lee}, H., \& {Lee}, H.~M. 2002, ApJ, 567, 922

\bibitem[{{Dijkstra} {et~al.}(2005){Dijkstra}, {Haiman}, \&
  {Spaans}}]{dijkstra05a}
{Dijkstra}, M., {Haiman}, Z., \& {Spaans}, M. 2005, astro-ph/0510407

\bibitem[{{Eddington}(1926)}]{eddington26}
{Eddington}, A.~S. 1926, The internal constitution of the stars (London:
  Cambridge University Press)

\bibitem[{{Harrington}(1973)}]{harrington73}
{Harrington}, J.~P. 1973, \mnras, 162, 43

\bibitem[{{Neufeld}(1990)}]{neufeld90}
{Neufeld}, D.~A. 1990, ApJ, 350, 216

\bibitem[{{Norman}(2004)}]{norman}
{Norman}, M.~L. 2004, in Adaptive Mesh Refinement - Theory and Applications,
  Eds. T. Plewa, T. Linde and V. G. Weirs, Springer Lecture Notes in
  Computational Science and Engineering

\bibitem[{{Phillips} \& {Meszaros}(1986)}]{phillips_meszaros86}
{Phillips}, K.~C. \& {Meszaros}, P. 1986, \apj, 310, 284

\bibitem[{{Stenflo}(1980)}]{stenflo80}
{Stenflo}, J.~O. 1980, Astron. \& Astrophys., 84, 68

\bibitem[{{Tasitsiomi}(2005)}]{tasitsiomi05b}
{Tasitsiomi}, A. 2005, astro-ph/0510347

\bibitem[{{Unno}(1955)}]{unno55}
{Unno}, W. 1955, \pasj, 7, 81

\end{thebibliography}

\end{document}